\documentclass[runningheads]{cl2emult}

\usepackage{makeidx}  % allows index generation
\usepackage{graphicx} % standard LaTeX graphics tool
\usepackage{subeqnar} % subnumbers individual equations
\usepackage{multicol} % used for the two-column index
\usepackage{cropmark} % cropmarks for pages without
\usepackage{lnp}      % placeholder for figures
\makeindex            % used for the subject index
\begin{document}
\title*{X-ray Surveys of the obscured Universe}
\toctitle{X-ray Surveys of the obscured Universe}
% allows explicit linebreak for the table of content
%
%
\titlerunning{X-ray Surveys}
% allows abbreviation of title, if the full title is too long
% to fit in the running head
%
\author{G\"unther Hasinger\inst{1,2}}
\authorrunning{G\"unther Hasinger}
% if there are more than two authors,
% please abbreviate author list for running head
%
%
\institute{Astrophysikalisches Institut Potsdam, An der Sternwarte 16,
           14482 Potsdam
\and Universit\"at Potsdam, Am Neuen Palais 10, 14469 Potsdam}

\maketitle              % typesets the title of the contribution

\begin{abstract}
Deep X-ray surveys have shown that the cosmic X-ray background (XRB) is largely
due to accretion onto supermassive black holes, integrated over cosmic time. 
However, the characteristic hard spectrum of the XRB can only be explained
if most AGN spectra are heavily absorbed. The absorbed AGN will suffer severe 
extinction and therefore, unlike classical QSOs, will not be prominent at 
optical wavelengths. Most of the accretion power is being absorbed by gas
and dust and will have to be reradiated in the FIR/sub-mm band. 
AGN could therefore contribute a substantial fraction 
to the recently discovered cosmic FIR/sub-mm background.
Here it is shown that a number of high-redshift absorbed X-ray sources 
selected in the {\it ROSAT} Deep Survey of the Lockman Hole
have broad-band spectral energy distributions very similar to the
local ULIRG NGC6240, lending additional support to the background models 
for the obscured universe.

\end{abstract}

\section{Echos from the past}

All cosmic processes like the Big Bang, the formation and evolution of 
galaxies, the creation and the growth of massive Black Holes 
and the heating of the Universe due to large-scale clustering will 
imprint characteristic radiation patterns on the electromagnetic spectrum
of the universe. The structure of the Universe at the epoch of decoupling is 
frozen into the 2.7 K Cosmic Microwave Background (CMB). 
Star and galaxy light, produced after the Dark Ages by thermonuclear fusion, 
is mainly confined to the near-infrared, optical and ultraviolet spectral 
bands. Active black holes in galactic nuclei (AGN), which accrete matter and 
efficiently convert gravitational energy into radiation, typically shine in a 
very broad energy band, from radio to gamma wavelengths. The role of dust is 
also very important: it absorbs
optical, UV and X-ray light, is warmed up and re-radiates the energy
in the far-infrared band. The extragalactic background radiation, i.e. the
total cosmic energy density spectrum, therefore gives a fossil record of 
all radiation processes in the universe, integrated over cosmic time. 
 
\begin{figure}
\centering
\centerline{
\includegraphics[width=.8\textwidth,clip=]{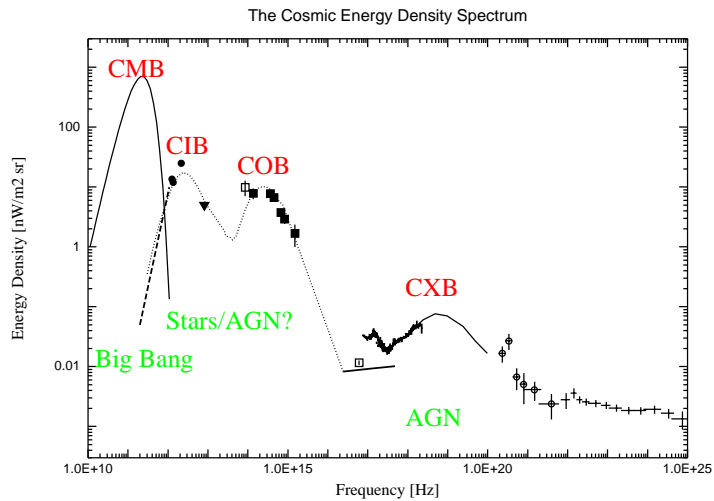}}
\caption[]{The cosmic energy density spectrum from radio waves to high energy
gamma rays in a $\nu I_\nu$ representation, where a horizontal line corresponds
to equal radiation power per decade of energy. The FIR to UV data
are from (left to right) \cite{pug96,fix98,hau98,vlk98,poz98,arm94}. The 
dotted line shows one of the models in \cite{dwe98} extrapolated from 
the visual to the EUV range. The X-ray/gamma-ray data are from 
\cite{has93,war98,miy98,gru92,kap96,sre98}.}
\label{ECHOS}
\end{figure}

In the recent years, the extragalactic background spectrum has been determined
with relatively high precision over a very broad range of the electromagnetic 
spectrum, apart from some inaccessible regions. Figure \ref{ECHOS} shows a 
compilation of most recent 
determinations of the cosmic energy density spectrum from radio waves to 
high-energy gamma rays -- the "Echos from the Past". 
Apart from the CMB peak, which clearly dominates the energy budget of the 
universe, three distinct components can be identified in the spectral energy 
distribution: the Cosmic Far-Infrared Background (CIB), the Cosmic Optical/UV 
Background (COB) and the Cosmic X-ray/Gamma-ray Background (CXB).

The cosmic X-ray background (CXB) radiation  was
the first extragalactic background emission to be announced\cite{gia62}. 
In the seventies, X--ray satellites like {\it UHURU}, {\it ARIEL V} 
and {\it HEAO--1}, which were able to scan the whole sky, produced high-quality
XRB maps revealing a large degree of isotropy which immediately led to the 
conclusion that the origin of the XRB has to be mainly extragalactic and might 
therefore be of cosmological interest. Gamma-ray satellites like {\it SAS-3}, 
{\it COS-B}
and finally the Compton Gamma Ray Observatory extended the background 
measurements to the highest-energy gamma rays
 (see \cite{has96} for a review). The energy spectrum of the CXB
has a characteristic hump at about 30 keV and a long tail towards higher 
energies. At energies below 1 keV diffuse galactic emission starts to dominate
the sky. At 1/4 keV a large fraction of the background determined by shadowing
experiments has been resolved by deep {\it ROSAT} surveys\cite{war98,has93}. 

In the sub-mm to UV range the extragalactic background light is 
completely swamped by other background components, i.e. the CMB, the 
interplanetary
dust emission, the interstellar dust emission, scattering from the 
interplanetary dust (zodiacal light), galactic starlight and scattering by
interstellar dust. 
In the far-infrared range different groups have 
recently detected the long-sought residual cosmic far-infrared background (CIB)
signal\cite{pug96,hau98,fix98} by carefully modelling the other components
in the high-frequency tail of the Cosmic Microwave Background (see fig. \ref{ECHOS}). 

In the near-infrared to 
UV band, on the other hand, the Hubble Space Telescope Deep Field images are so 
sensitive, 
that one can safely assume that most of the Cosmic Optical Background (COB) has
been resolved into discrete sources\cite{poz98}. 
In the mid-infrared range the bright zodiacal light does not allow a 
direct detection of the extragalactic background light and one has to resort to
galaxy population synthesis models and relatively uncertain upper 
limits\cite{dwe98}.
At higher frequencies, in the extreme ultraviolet range, the interstellar 
absorption basically inhibits the detection of the extragalactic light and
the background level can only be estimated by interpolating between the UV and
the soft X-ray measurements. 
  
The extragalactic background light seems to be dominated by
two distinct humps, one in the optical/near-infrared range (COB), which could 
very
well be produced by the red-shifted star light in all of the distant galaxies,
and another one in the FIR/sub-mm range (CIB), which is very likely 
produced by dust emission in distant galaxies. The fact that the dust peak
contains at least as much or even more energy than the stellar peak 
indicates that a large amount of light in the early universe had to be 
absorbed by dust. However, what fraction of the dust is heated by starlight 
or by AGN accretion power remains to be determined. 

\section{Deep X-ray surveys}

The X-ray range is one of the few regions of the electromagnetic spectrum,
where on one hand the sky emission is dominated by the extragalactic 
background and on the other hand modern imaging telescopes are sensitive
enough to resolve a substantial fraction of the background into discrete
sources. The aim of deep X-ray surveys is to resolve as much of the background
as possible, then to obtain reliable optical identifications and redshifts
of the X-ray sources and finally for all classes of objects to determine 
luminosity functions and their cosmological evolution (see e.g. \cite{miy00}). 
Since the emitted light, integrated 
over cosmic time can be compared directly to the total X-ray background,
important constraints on the cosmological evolution and possible
residual diffuse intergalactic emission can be obtained.

\begin{figure}
%\centering
\parbox{7cm}{\includegraphics[width=.5\textwidth,clip=]{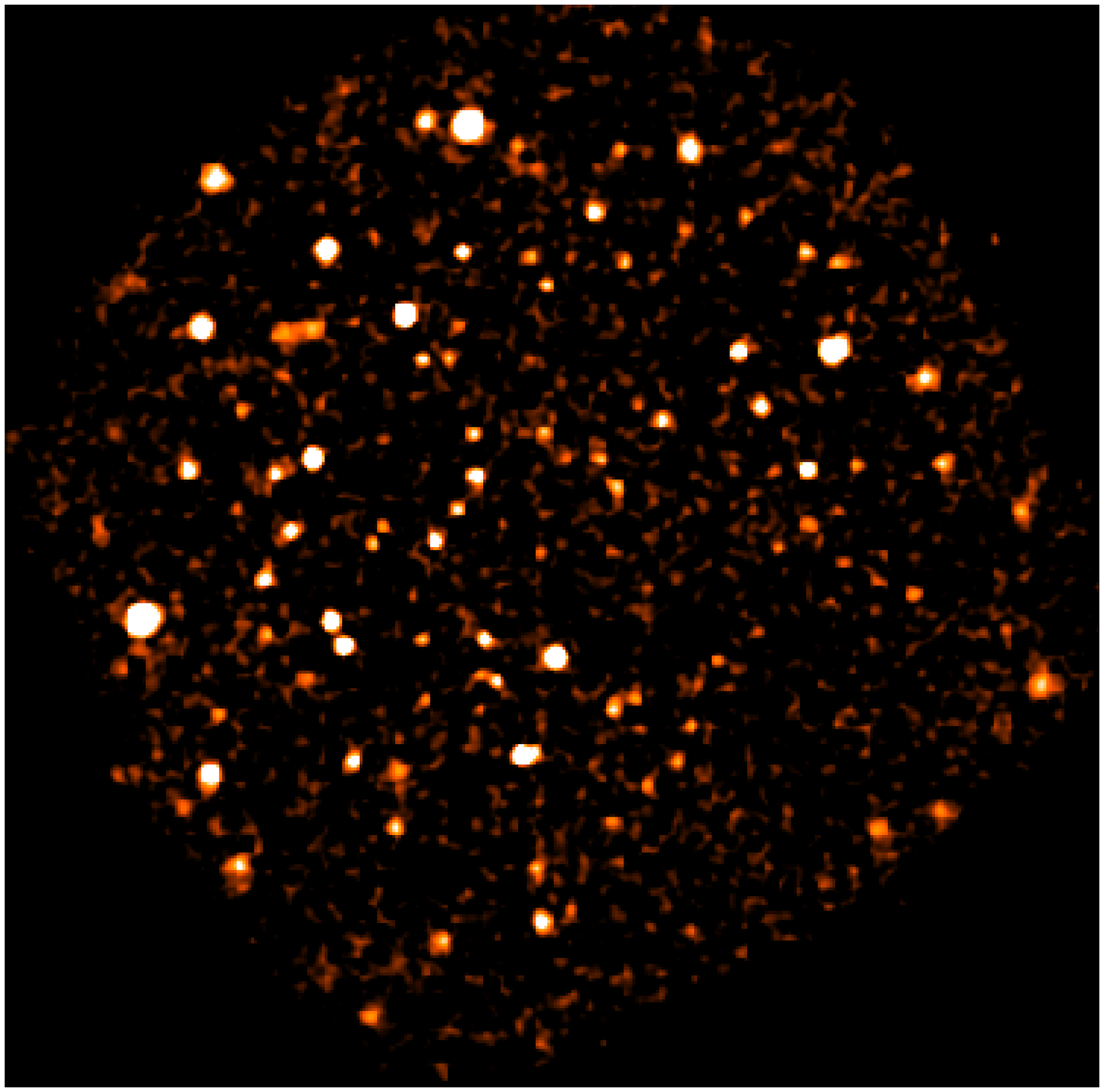}}
\ \hspace{-1.3cm} \
\parbox{7cm}{\includegraphics[width=.5\textwidth,clip=]{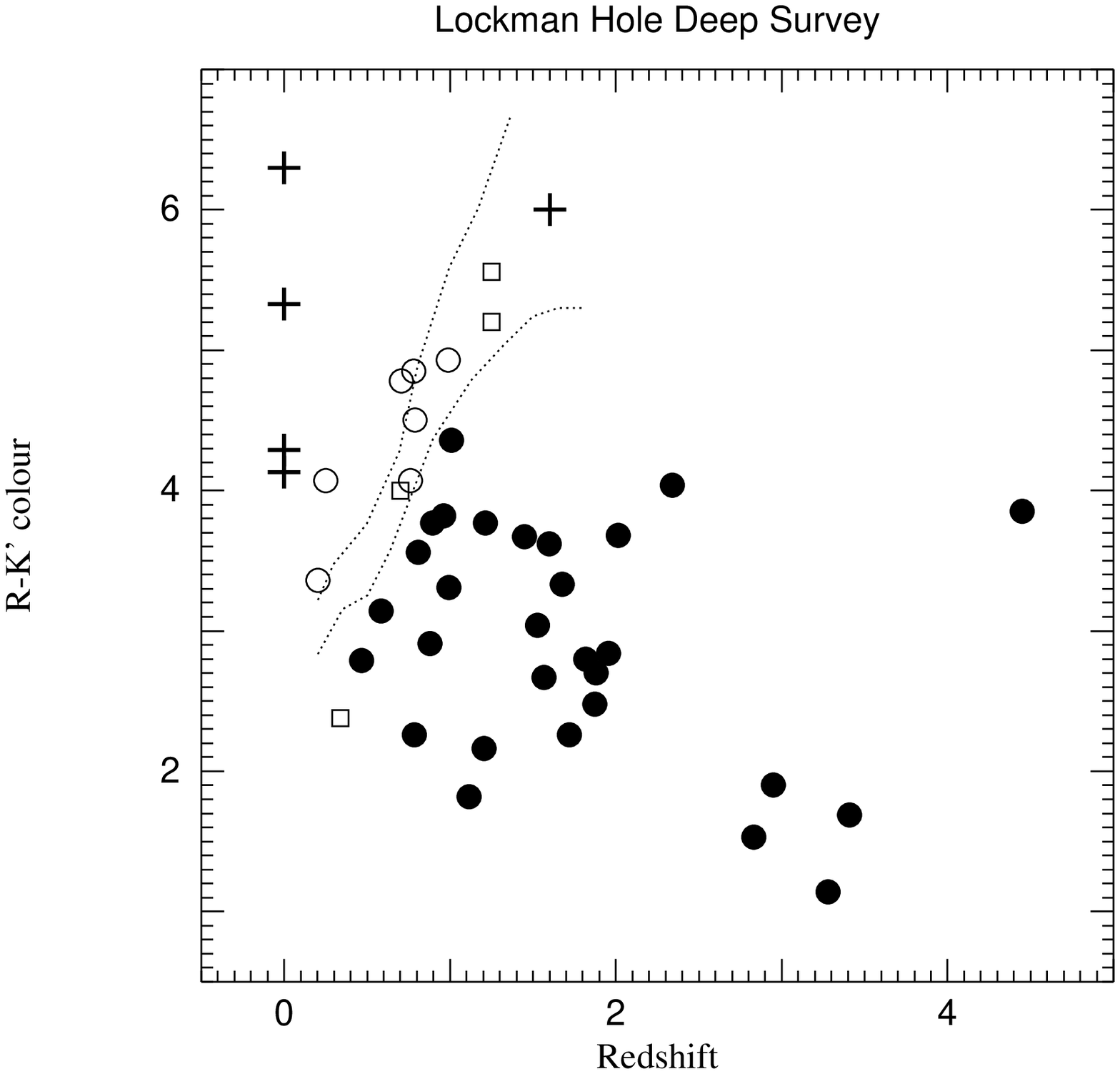}}
\caption[]{{\bf Left:}The {\it ROSAT} Deep HRI Survey in the Lockman 
Hole\cite{has98}; the size of the image is about 35 arcmin across.
{\bf Right: } R-K colours against redshift for X-ray sources in the Lockman 
Hole marked with different symbols: filled circles are broad-line AGN, 
open circles are narrow-line AGN (Sy 2), open squares are clusters and groups 
of galaxies. Plus signs are optical/NIR counterparts without spectroscopic 
identifications, for the ROSAT source 14Z with a photometric redshift 
$z_{ph} \approx 1.6$. The dotted lines correspond to unevolved spectral models 
for E (upper) and S$_{\rm b}$ (lower) galaxies\cite{bru93}.}
\label{HRI}
\end{figure}

A large number of deep, medium-deep and shallower surveys have been performed
with {\it ROSAT} (see \cite{has99} for a review).
The {\it ROSAT} Deep Surveys in the Lockman Hole (see \ref{HRI}a)
have so far been the deepest 
X-ray observations ever performed (until the deep {\it Chandra} and {\it XMM}
surveys will be completed). They reach a source density of 
$\sim 1000~{\rm sources~deg}^{-2}$ at the faintest flux limit,
where 70--80\% of the soft X--ray background
has been resolved into discrete sources\cite{has98}. 

Deep R-band images have been obtained at the Keck telescope. 
In the Lockman Hole Deep survey most
optical counterparts have magnitudes in the range R=18--23. 
With the excellent HRI positions typically
only one or two counterparts are within the X--ray error circle. 
Spectra of the 
optical counterparts of the X-ray sources were mostly obtained with the 
Keck telescope. Currently 85 out of 91 X--ray sources are spectroscopically 
identified. About 80\% of those turned out to be active galactic 
nuclei\cite{schm98}, among them the highest redshift X--ray selected QSO at
z=4.45\cite{schn98}. Most AGN are luminous QSOs and Sy1 galaxies with broad 
emission lines. However about 16\% show only narrow emission lines and are 
classified as Seyfert 2 galaxies\cite{has99}.

There is a group of X-ray sources in the Lockman Hole, which 
have optical counterparts fainter than R=24, for which optical spectroscopy
was not possible, even with the Keck telescope. For these one has to resort 
to NIR photometry and photometric redshifts. 
It turns out that the spectroscopically unidentified objects
typically have very red R-K colours\cite{has99,leh99}.  
Figure \ref{HRI}b shows a correlation diagram between the R-K colours
and redshifts for different types of objects identified in the Lockman Hole
in comparison with normal galaxy colours predicted from evolutionary models. 
The colours of type-1 AGN are relatively blue and scatter widely,
while type-2 AGN and cluster galaxies show on average much redder colours,
correlating well with redshift
in the region expected for normal galaxies. This indicates that the
optical colours of type-2 AGN are dominated by the light from the host
galaxy, probably because the optical nucleus is obscured. 
A population of heavily obscured AGN is thus already 
showing up in the soft X-ray band. Based on R-K colours 
and photometric redshifts it is likely that the as yet unidentified sources 
are dust shrouded AGN in the redshift range 1-2. 

There is, however, an interesting dichotomy in the 
redshift distribution of the sources. Unobscured objects (solid circles) are 
found to $z\approx4.5$, while the obscured objects (open circles, pluses) are
restricted to $z<2$. There is a ''negative K-correction effect''
for obscured objects at soft X-rays: at higher redshifts it  
is easier to detect obscured X-ray sources, because their unobscured 
rest-frame hard X-ray emission is shifted into the observed soft X-ray band.  
The fact that we do not find obscured objects at high redshifts could indicate 
a dependence of obscuration on intrisic source luminosity. High luminosity 
X-ray sources to some degree could be able to ''clean out'' their environment 
from obscuring material.  

\section{Hard X--ray Population Synthesis Models}

The X--ray background has a significantly harder spectrum than that of
local unobscured AGN. This led to the assumption that
a large fraction of the background flux is due to obscured AGN, as
originally proposed by Setti and Woltjer \cite{swo89}. Models following the
unified AGN schemes, assuming an appropriate mixture of absorbed and
unabsorbed AGN spectra folded with cosmological AGN evolution models, can 
in principle explain the shape of the background spectrum over the whole 
X--ray band (e.g. \cite{mat94,mad94,com95}), but gave problems with a number 
of new observational constraints\cite{gil99}. Apart from the AGN cosmological 
evolution\cite{miy00}, the distribution of absorption column densities 
among different types of AGN and as a function of redshift is one of the major 
uncertainties. The standard X--ray background population 
synthesis models (e.g. \cite{com95}) assume that the absorption distribution, 
which has been determined observationally only for local Seyfert galaxies,
is independent of X--ray luminosity and redshift. In particular, the fraction 
of type-2 QSOs, obscured high-luminosity X-ray sources, should be as high 
as that of Seyfert-2 galaxies. This assumption, however,
has been proven wrong by 
recent observations (see e.g. fig. \ref{HRI}b, \cite{miy00,gil99}).
The current X-ray background models therefore still have rather limited
predictive power. 

A significant improvement in our understanding of the hard X-ray background
is to be expected from optical identifications of complete samples of sources
selected at faint fluxes in the hard X--ray band.
Deep Surveys performed with {\it ASCA}\cite{geo97,oga98} and 
{\it BeppoSAX}\cite{fio99} have recently resolved about 30\% of the 
harder X-ray background, but due to very limited angular resolution
their optical identification is tedious and in the case of faint 
optical counterparts almost impossible.
Upcoming deep surveys with the {\it Chandra} observatory ({\it AXAFi}) and 
{\it XMM}
with very high sensitivity and good positional accuracy in the hard band
together with optical identifications from the VLT and the Keck telescopes
are expected to yield a solid statistical basis to disentangle these
various effects and lead to a new, unambiguous population synthesis for the
X--ray background.

An immediate consequence of the obscured background models is that the radiation
produced by accretion processes in AGN emerges completely
unabsorbed only at energies well above 10 keV, thus producing the observed
maximum of the XRB energy density at $\sim 30$ keV. Comparison between
the background energy density at 30 keV and at 1 keV leads to the
suggestion that most (80--90\%) of the accretion power in the universe
might be absorbed, implying a very large solid angle of the obscuring
material as seen from the central source. Fabian et al. \cite{fab98} suggest
that circumnuclear starburst regions are responsible for the large covering
factor. They may be both triggering and obscuring most of the nuclear
activity. Recently Fabian and Iwasawa \cite{fab99} have shown that these
AGN background synthesis models can
explain the mass distribution of dark remnant black holes in the
centers of nearby galaxies by conventional accretion which is
largely hidden by obscuration.

\section{Predictions for other wavebands}

The obscured background synthesis models has important consequences
for the current attempts to understand black hole and galaxy formation
and evolution. The absorbed AGN will suffer severe extinction and
therefore, unlike classical QSOs, would not be prominent at optical
wavelengths. The light of the optical counterparts of distant absorbed
X-ray sources should therefore be dominated by the host galaxy, which indeed
seems to be the case for a significant number of faint X-ray sources (see
fig \ref{HRI}b).  

If most of the
accretion power is being absorbed by gas and dust, it will have to be
reradiated in the FIR range and be redshifted into the sub--mm
band. AGN could therefore contribute a substantial fraction to the
cosmic FIR/sub--mm background which has already partly
been resolved by deep SCUBA surveys. The background population
synthesis models have recently been used by Almaini et al. \cite{alm99}
to predict the AGN contribution to the sub--mm background
and source counts. Depending on the assumptions about cosmology and
AGN space density at high redshifts 
they predict that a substantial fraction of the sub--mm source counts
at the current SCUBA flux limit could be associated with active galactic
nuclei. Interestingly, the first optical identifications of SCUBA sources
indicate a significant AGN contribution\cite{bar99}.

Another, largely independent line of arguments leads to the conclusion
that accretion processes may produce an important contribution to the
extragalactic background light. Dynamical studies\cite{mag98}
come to the conclusion that massive dark objects, most
likely dormant black holes, are ubiquitous in nearby galaxies. There
is a correlation between the black hole mass and the bulge mass of
a galaxy: $M_{BH} \approx 6 \times 10^{-3} M_{Bulge}$. Since gravitational
energy release through standard accretion of matter onto a black hole is
producing radiation about 100 times more efficiently than the thermonuclear
fusion processes in stars, the total amount of light produced by accretion
in the universe should be of the same order of magnitude as that produced by
stars. A more detailed treatment following this argument comes to the
conclusion that the bolometric AGN contribution should be about 1/5 of the 
stellar light in the universe\cite{fab99}.

\section{NGC 6240 -- representative of the obscured universe?}

\begin{figure}
\centering
\centerline{
\includegraphics[width=.7\textwidth,clip=]{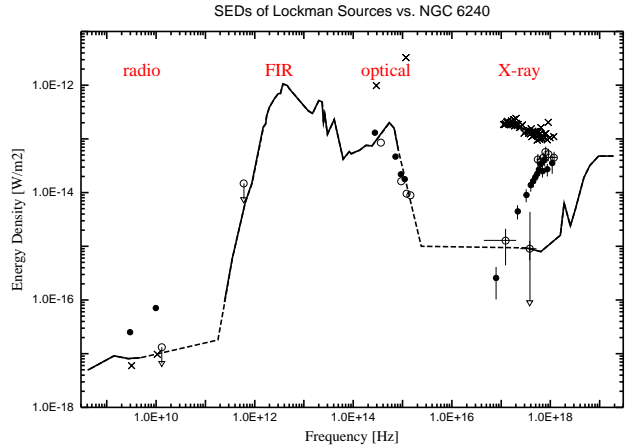}}
\caption[]{Spectral energy distribution of the ultraluminous IRAS 
galaxy (ULIRG) NGC 6240 from radio to hard X-ray frequencies (thick
solid line, dashed in unobserved portions)\cite{NED,kla97,vig99},
compared to the spectral energy distribution of three high-redshift
X-ray sources from the {\it ROSAT} Deep Survey\cite{schm98,leh99}: 12A 
(filled circles), 14Z (open circles) and 32A (crosses). The 
latter SEDs have been normalised to the redshift of NGC 6240 (see text).}
\label{SED}
\end{figure}

The ultraluminous IRAS galaxy NGC 6420 is a starburst galaxy with
a double nucleus\cite{raf97} with some evidence for a weak active nucleus.
Its position in the mid-infrared diagnostic diagram based on ISO spectroscopy
\cite{lut96} indicates that its bolometric luminosity, which is dominated
by the far-infrared dust emission, should be 
mainly heated by star formation processes. Recently, however, {\it BeppoSAX}
hard X-ray spectroscopy revealed an underlying very luminous, though 
heavily obscured active galactic nucleus, which could well be responsible
for the major fraction of the bolometric luminosity. Figure \ref{SED}
shows the spectral energy distribution (SED) of NGC 6240 in a $\nu F_\nu$ 
representation from the radio to the hard X-ray band. The SED peaks in the 
far-infrared dust band and shows the mid-infrared policyclic aromatic 
hydrocarbonic emission bands as well as the synchrotron radio emission
typical for starburst galaxies. The 
optical starlight peak is lower than the dust emission peak - similar to 
the overall extragalactic energy distribution. In the hard X-ray range
the nonthermal AGN continuum emission is severely absorbed below  
15 keV and strong reprocessed iron line is seen. The UV to soft X-ray 
nuclear continuum is completely obscured.   

The SEDs of three AGN from the 
{\it ROSAT} deep survey in the Lockman Hole, which have been 
observed over a broad wavelength range from radio to X-rays have been  
superposed on this diagram. These are (see \cite{schm98,leh99}):
32A, a QSO/Sy1 at $z$=1.33 (crosses), 12A a Sy2 at $z$=0.99 (filled circles)
and 14Z, a very red object, probably a Sy2 galaxy, with 
$z_{ph}\approx1.6$\cite{poz99}. The radio
detections and upper limits at 20cm and 6cm are from \cite{rui97} and 
\cite{cil00}. The upper limit in the millimeter-range for 14Z has been 
obtained with IRAM\cite{ber00}. 
In the X-ray range the {\it ROSAT PSPC} spectra of these three sources have 
been fitted with a power law with galactic or intrinsic absorption. 
For a comparison on the same absolute 
scale the Lockman SEDs have been shifted to the 
redshift of NGC 6240 and scaled by the square of their luminosity
distances. 

The two absorbed {\it ROSAT} sources (12A and 14Z) have SEDs 
surprisingly similar to NGC 6240. The galaxy 12A may have a relatively 
stronger starburst component, indicated by its higher radio luminosity. 
The main difference lies in the X-ray range, where the luminosity at 
hard X-ray energies is very similar for all 3 sources, while the intrinsic 
absorption is clearly different: $N_H$ is about $10^{24}cm^{-2}$ for 
NGC6420, $3\cdot10^{22}cm^{-2}$ for 14Z and $10^{22}cm^{-2}$ for 12A. 
In contrast, the SED for the radio-quiet QSO 32A, while
fitting relatively well to the other sources in the radio band, is 
clearly dominated by the unobscured light of the active nucleus in
the optical and X-ray range.

This figure is a nice illustration of the obscured AGN model for the 
X-ray background and its potential relevance to the FIR/sub-mm
background, since it shows that X-ray-selected high-redshift AGN can
indeed have SEDs similar to local ULIRGS. 
Unfortunately there are no observations yet, which are sensitive 
enough both in the X-ray band and in the sub-mm band to obtain a better
quantitative estimate of the fraction of the far-infrared background
produced by AGN. Future joint sub--mm/X--ray deep surveys will therefore be 
a very powerful tool to disentangle the different processes dominating the 
universe at high redshift.

{\small {\it Acknowledgements:}
I thank my collaborators in various projects, in particular F. Bertoldi,
P. Ciliegi, R. Giacconi, I. Lehmann, M. Schmidt, T. Soifer, D. Thompson,
J. Tr\"umper, G. Wilson and G. Zamorani for the permission to show some 
material in advance of 
publication. This research has made use of the NASA/IPAC Extragalactic
Database (NED) which is operated by the Jet Propulsion Laboratory, California 
Institute of Technology, under contract with the National Aeronautics and 
Space Administration. This work has been supported in part by DLR grant 
50~OR~9908~0.}

%INDEX%%%%%%%%%%%%%%%%%%%%%%%%%%%%%%%%%%%%%%%%%%%%%%%%%%%%%%%%%%%%%%%
\clearpage
\addcontentsline{toc}{section}{Index}
\flushbottom
\printindex
%%%%%%%%%%%%%%%%%%%%%%%%%%%%%%%%%%%%%%%%%%%%%%%%%%%%%%%%%%%%%%%%%%%%%


\begin{thebibliography}{7}
%
\addcontentsline{toc}{section}{References}

\bibitem{alm99}Almaini O., Lawrence A., Boyle B., 1999, MNRAS 305, L59

\bibitem{arm94} Armand C., et al., 1994, A\&A 284, 12

\bibitem{bar99} Barger A.J., et al., 1999, AJ117, 2656

\bibitem{ber00} Bertoldi F., et al., 2000 (in prep.)

\bibitem{bru93} Bruzual A.G. \& Charlot S., 1993, ApJ 405, 538 

\bibitem{cen99} Cen R., Ostriker J., 1999, ApJ 514, 1

\bibitem{cil00} Ciliegi P., et al., 2000 (in prep.)

\bibitem{com95} Comastri A., et al., A\&A 296, 1

\bibitem{dwe98} Dwek E., et al., 1998, ApJ 508, 106

\bibitem{fab98}Fabian A.C. et al., 1998, MNRAS 297, L11

\bibitem{fab99}Fabian A.C., Iwasawa K., 1999, MNRAS 303, L34

\bibitem{fio99}Fiore F. et al., 1999, MNRAS 306, L55

\bibitem{fix98} Fixsen D.J., et al., 1998, ApJ 508, 123

\bibitem{gia62} Giacconi, R. et al., 1962, Phys.Rev.Letters 9, 439

\bibitem{gil99}Gilli R., Risaliti G. \& Salvati M., 1999, A\&A 347, 424

\bibitem{geo97}Georgantopoulos I. et al., 1997, MNRAS 291, 203

\bibitem{gru92} Gruber D.E., in {\it The X--ray Background} X. Barcons \& A.C.
   Fabian eds, (Cambridge University Press), p.45 (1992).

\bibitem{has93} Hasinger G., et al., 1993, A\&A 295, 1 

\bibitem{has96} Hasinger G., 1996, A\&AS 120C, 607 

\bibitem{has98} Hasinger G., et al., 1998, A\&A 329, 482 

\bibitem{has99} Hasinger G. et al., 1999, astro-ph/9901103

\bibitem{hau98} Hauser M.G., et al., 1998, ApJ 508, 25 

\bibitem{leh99} Lehmann, I., et al., 1999, A\&A (in press), astro-ph/9911484

\bibitem{lut96} Lutz D., et al., 1996, A\&A 315, L137 

\bibitem{kap96} Kappadath, S.C., et al., 1996, A\&AS 120C, 619

\bibitem{kla97} Klaas U., et al., 1997, A\&A 325, L21 

\bibitem{mad94} Madau P. et al., 1994, MNRAS 283 1388

\bibitem{mag98}Magorrian J. et al., 1998, AJ 115, 2285

\bibitem{mat94} Matt G. \& Fabian A.C., 1994, MNRAS 267 187

\bibitem{miy98} Miyaji T. et al., 1998, A\&A 334, L13 

\bibitem{miy00} Miyaji T. et al., 2000, A\&A 353, 25 

\bibitem{oga98}Ogasaka Y. et al. 1998, AN 319, 43

\bibitem{poz98} Pozzetti L. et al., 1998, MNRAS 298, 1133 

\bibitem{poz99} Pozzetti L. 1999, priv. comm. 

\bibitem{pug96} Puget J.-L. et al., 1996, A\&A 308, L5

\bibitem{raf97} Rafanelli P., et al. 1997, A\&A 327, 901  

\bibitem{rui97} de Ruiter H. et al., 1997, A\&A 319, 7  

\bibitem{schm98}Schmidt M. et al. 1998, A\&A, 329, 495

\bibitem{schn98}Schneider D. et al., 1998, AJ 115, 1230

\bibitem{sno97} Snowden S.L., et al., 1997, ApJ 485, 125

\bibitem{swo89} Setti G. and Woltjer L. 1989, A\&A 224, L21

\bibitem{sre98} Sreekumar P., et al., 1998, ApJ 494, 523

\bibitem{teg97} Tegmark M., et al., 1997, ApJ 474, 1

\bibitem{vig99} Vignati P., et al., 1999, A\&A 349, 57 

\bibitem{vlk98} V\"olk H. 1998, priv. comm.

\bibitem{war98} Warwick R. S., Roberts T. P, 1998, AN 319, 59
 
\bibitem{zdz95} Zdziarski A.A., et al., 1995, ApJ 438, L63 
 
\bibitem{NED} NED 
\end{thebibliography}
\end{document}